\documentclass[twocolumn,tighten,onecolappendix]{aastex63}

\usepackage{amssymb,amsmath,amsthm}
\usepackage{graphicx}
\usepackage{mathrsfs}
\usepackage{booktabs}
\usepackage{multirow}
\usepackage{empheq}
\usepackage{verbatim}
\usepackage{mathtools}
\usepackage{hyperref}
\usepackage{xcolor}
\usepackage{xspace}
\usepackage[makeroom]{cancel}
\usepackage[acronym]{glossaries}

\newcommand{\harvard}{Harvard College, Harvard University, Cambridge, MA 02138, USA}
\newcommand{\cfa}{Center for Astrophysics $|$ Harvard \& Smithsonian, 60 Garden Street, Cambridge, MA 02138, USA}
\newcommand{\bhi}{Black Hole Initiative, Harvard University, 20 Garden Street, Cambridge, MA 02138, USA}

\makeatletter
\def\frontmatter@title@above{}
\makeatother
\journalinfo{Accepted for publication in The Astronomical Journal}

\begin{document}

% \submitted{}

\title{Modeling atmospheric phase corruptions in high-frequency VLBI using Gaussian processes}

\correspondingauthor{Uri~Rolls}
\email{urirolls@college.harvard.edu}

\author{Uri~Rolls}
\affil{\harvard}
\author[0000-0002-5278-9221]{Dominic W. Pesce}
\affil{\cfa}\affil{\bhi}
\author[0000-0003-3826-5648]{Paul Tiede}
\affil{\cfa}\affil{\bhi}
\author[0000-0002-9030-642X]{Lindy Blackburn}
\affil{\cfa}\affil{\bhi}
\author[0000-0001-8242-4373]{Iniyan Natarajan}
\affil{\cfa}\affil{\bhi}
\author[0000-0002-9031-0904]{Sheperd S. Doeleman}
\affil{\cfa}\affil{\bhi}

\begin{abstract}
Using very long baseline interferometry (VLBI) observations at (sub)millimeter wavelengths, the Event Horizon Telescope (EHT) currently achieves the finest angular resolution of any astronomical facility, necessary for imaging the horizon-scale structure around supermassive black holes.  A significant calibration challenge for high-frequency VLBI stems from rapid variations in the atmospheric water vapor content above each telescope in the array, which induce corresponding fluctuations in the phase of the correlated signal that limit the coherent integration time and thus the achievable sensitivity.  In this paper, we introduce a model that describes station-based phase corruptions jointly with a parameterization for the source structure.  We adopt a Gaussian Process (GP) prescription for the time evolution of these phase corruptions, which provides sufficient flexibility to capture even highly erratic phase behavior. The use of GPs permits the application of a Kalman filtering algorithm for numerical marginalization of these phase corruptions, which permits efficient exploration of the remaining parameter space.  Our model also removes the need to specify an arbitrary ``reference station'' during calibration, instead establishing a global phase zeropoint by enforcing the GPs at all stations to have fixed mean and finite variance.  We validate our method using a real EHT observation of the blazar 3C 279, demonstrating that our approach yields calibration solutions that are consistent with those determined by the EHT Collaboration.  The model presented here can be straightforwardly extended to incorporate frequency-dependent phase behavior, such as is relevant for the frequency phase transfer calibration technique.
\end{abstract}

\section{Introduction} \label{sec:Introduction}

A common issue faced in the calibration of radio interferometric data is the presence of substantial and time-variable corruptions in the primary measurement quantity, the complex visibility. For high-frequency very long baseline interferometry (VLBI) observations in particular, such as those conducted by the Event Horizon Telescope (EHT), these corruptions are especially significant and rapid in the phase of the complex visibility \citep{EHT_M87_Paper3}. These phase variations -- primarily introduced by fluctuations in the distribution of water vapor in the troposphere above each station participating in the array -- limit the coherence time of observations and 
thus the length of time over which VLBI signals can be coherently integrated.  Effective mitigation of these time-variable phase corruptions is essential for achieving imaging capabilities with VLBI arrays.

At observing frequencies above $\sim$100\,GHz, traditional ``phase referencing'' calibration \citep[e.g.,][]{Beasley_1995} becomes impractical, because the atmospheric variability timescales are too short and the typical separation between any given science target and a sufficiently bright calibrator is too large.  Instead, ``self-calibration'' techniques are employed that determine phase solutions using the science target itself \citep[e.g.,][]{Readhead_1978,Pearson_1984}.  Such techniques take advantage of the station-based origin of the dominant phase variations to decouple the corruptions from the underlying source visibility phase structure.  Self-calibration as applied to EHT data has to date followed an algorithmic inverse-modeling approach, whereby visibility phase measurements on baselines to a sensitive ``reference station'' are treated as station phases and removed in a manner that stabilizes the time variability while preserving phase closure \citep[e.g.,][]{EHT_M87_Paper3,Blackburn_2019}.

In this paper, we present an alternative approach for VLBI phase calibration that forward-models the phase contributions from each station alongside a prescription for the source visibility phase.  We model the time-variable phases as Gaussian Processes (GPs) with a parameterized covariance structure that is free to vary across stations.  Our approach is conceptually aligned with a number of other recent advances in forward-modeling techniques for VLBI, particularly those that jointly model corrupting effects alongside the source structure within a Bayesian framework \citep[e.g.,][]{Natarajan_2017,Natarajan_2020,Resolve,Themis,DMC,Tiede_2022}, as well as modeling of tropospheric turbulence as a Gaussian process for the purpose of data simulation \citep{Blecher_2017,Natarajan_2022}.  The primary difference we advance here is in calibrating data that are further ``upstream'' than most previous approaches; i.e., our calibration model does not require any pre-processing to stabilize nonlinear visibility phase variations in time.

This paper is structured as follows. In \autoref{sec:Model}, we detail our phase model and describe its implementation using a Kalman filtering algorithm. In \autoref{sec:ApplicationSynthetic} we validate our method using synthetic VLBI datasets, and in \autoref{sec:ApplicationReal} we apply our approach to EHT observations and compare our results with the original EHT calibration.  In \autoref{sec:Discussion}, we present a discussion of our model, and we summarize and conclude in \autoref{sec:Summary}.

Throughout this paper, we adopt the notation for Gaussian distributions used in \citet{MatrixCookbook}, where a normally-distributed vector $\boldsymbol{x}$ with mean $\boldsymbol{\mu}$ and covariance $\boldsymbol{\Sigma}$ is denoted as
\begin{equation}
    \boldsymbol{x} \sim \mathcal{N}\left( \boldsymbol{\mu}, \boldsymbol{\Sigma} \right) .
\end{equation}
When necessary, we explicitly indicate the random variable in the subscript, e.g., $\mathcal{N}_{\boldsymbol{x}}\left( \boldsymbol{\mu}, \boldsymbol{\Sigma} \right)$, to avoid ambiguity.

\section{Model description} \label{sec:Model}

An interferometric baseline between stations $i$ and $j$ is sensitive to the complex visibility $V_{ij}$ of the on-sky source structure \citep{TMS}, which is related to the complex correlation of the electric fields $E_i$ and $E_j$ incident on the two stations,

\begin{equation}
V_{ij} \propto \left\langle E_i E_j^* \right\rangle . \label{eqn:Visibility}
\end{equation}

\noindent Here, the asterisk denotes complex conjugation, and the proportionality constant depends on the sensitivities of the participating stations \citep[see, e.g.,][]{Blackburn_2019}.  In this paper, we work exclusively with the phase of the complex visibility,

\begin{equation}
\phi_{ij} = \text{arg}(V_{ij}) .
\end{equation}

\subsection{Phase model} \label{sec:Instrument}

The relationship in \autoref{eqn:Visibility} is complicated in practice by the presence of corrupting effects in the data, the most severe of which are typically station-based and known as ``complex gain'' corruptions.
For high-frequency VLBI, the gain phase introduced by variations in the amount of tropospheric water vapor can dominate the signal, causing the visibility phase to vary by more than a radian on timescales of a few to tens of seconds \citep[e.g.,][]{EHT_M87_Paper2}.

Our basic phase model is given by

\begin{equation}
\hat{\phi}_{ij}(t) = \phi_{ij} + \theta_i(t) - \theta_j(t) + \varepsilon_{ij}(t) , \label{eqn:Model}
\end{equation}

\noindent where $\phi_{ij}$ represents the true source visibility phase that would be observed in the absence of gain corruptions, $\hat{\phi}_{ij}(t)$ is the measured visibility phase on baseline $ij$ at time $t$, $\theta_i(t)$ and $\theta_j(t)$ are the gain phases at stations $i$ and $j$, and $\varepsilon_{ij}(t)$ captures the statistical uncertainty (or ``thermal noise'') in the phase measurement $\hat{\phi}_{ij}(t)$.  We assume that the thermal noise contribution is Gaussian-distributed with variance $\sigma_{ij}^2(t)$,

\begin{equation}
\varepsilon_{ij}(t) \sim \mathcal{N}(0,\sigma_{ij}^2(t)) , \label{eqn:ThermalNoise}
\end{equation}

\noindent and that they are independent across stations and timestamps.  This assumption only strictly holds in the limit of high signal-to-noise ratio, but it remains approximately valid even down to signal-to-noise ratios as low as $\sim$2 \citep{Blackburn_2020}.

We model the time-dependent gain phases $\theta_i(t)$ and $\theta_j(t)$ as GPs. Specifically, the time series of gain phases at each station is assumed to follow a multivariate normal distribution with a parameterized covariance structure,

\begin{equation}
\begin{pmatrix}
\theta_i(t_1) \\
\theta_i(t_2) \\
\vdots \\
\theta_i(t_T)
\end{pmatrix} \equiv \boldsymbol{\theta}_i \sim \mathcal{N}(\boldsymbol{\mu}_i, \boldsymbol{\Sigma}_i) , \label{eqn:GPphases}
\end{equation}

\noindent where $\boldsymbol{\mu}_i$ is a vector representing the mean phase at each timestamp and $\boldsymbol{\Sigma}_i$ is the covariance matrix describing the gain phase structure for station $i$.  For the analyses presented in this paper, all elements of $\boldsymbol{\mu}_i$ are fixed to be zero.  The covariance structure of each GP is described by a kernel function, which can in principle take many forms.  For the analyses presented in this paper, we use a Mat\'ern-1/2 kernel, such that the elements of the covariance matrix are given by

\begin{equation}
( \Sigma_{i} )_{mn} = \sigma_i^2 \exp\left( -\frac{|t_m - t_n|}{\tau_i} \right) . \label{eqn:Matern12}
\end{equation}

\noindent Here, $\tau_i$ sets the variation timescale and $\sigma_i$ sets the magnitude of the phase fluctuations.  In \autoref{app:Kernel}, we assess the sensitivity of our measurements to the particular choice of kernel function.

For an array containing $N$ stations and making visibility phase measurements on each of $T$ timestamps, our model contains the following free parameters:
\begin{itemize}
    \item There are $N (N-1)/2$ visibility phase parameters $\phi_{ij}$, which we assume to be constant in time.
    \item There are $N T$ gain phase parameters $\theta_i(t_k)$, one for each station $i$ at each timestamp $t_k$.
    \item There are $2 N$ GP kernel parameters, a timescale $\tau_i$ and variance $\sigma_i^2$ for each station, which we assume to be constant in time.
\end{itemize}
\noindent We note that the source visibility phases $\phi_{ij}$ could in principle be inherited from a more general model of the source structure -- such as might be available during image reconstruction -- rather than being assumed constant.  However, for the analyses presented in this paper, we assume that the duration of the observation is sufficiently short that a constant-valued assumption for $\phi_{ij}$ is a reasonable approximation.

\subsection{Kalman filtering} \label{sec:Kalman}

For many applications, the gain phases $\theta_i$ are nuisance parameters that we model only for the sake of accessing the visibility phases $\phi_{ij}$, which are themselves the primary parameters of interest.  Because our phase model is linear in the gain phases (\autoref{eqn:Model}) and our measurement uncertainties are taken to be Gaussian (\autoref{eqn:ThermalNoise}), we can analytically marginalize over the $\theta_i$ parameters, yielding a posterior distribution for the remaining $\phi_{ij}$, $\tau_i$, and $\sigma_i$ parameters.  We derive the marginal posterior distribution for our model in \autoref{app:GainPhaseMarginalization}, but unfortunately this distribution is inefficient to work with in practice because it involves the construction and inversion of large (typical dimension ${\gg}10^3$) matrices.

Fortunately, our model is linear and Gaussian, and because we have restricted our covariance structure to the half-integer Mat\'ern family (see \autoref{eqn:Matern12}) it also admits a so-called ``state-space representation'' (see \citealt{Rozanov_1977, rozanov1982markov} and \autoref{app:KalmanFilters} for details).  This representation enables application of the efficient Kalman filtering algorithm to numerically solve and marginalize over the gain phases.  Kalman filtering is commonly employed for modeling time-series data \citep[e.g.,][]{Sarkka_2013}, where it takes advantage of the sequential nature of such data to iteratively refine a running solution using each data point consecutively.  I.e., given a solution derived from data points 1 through $k$, the Kalman filtering algorithm incorporates data point $k+1$ to update the existing solution rather than recomputing an entirely new one.  In this way, the complexity of the Kalman filtering algorithm grows only linearly with the number of data points, providing an efficient method for estimating the gain phases in our model.  We detail our implementation of the Kalman filtering algorithm in \autoref{app:KalmanFilters}.

\subsection{Parameter space exploration} \label{sec:ParameterExploration}

In our framework, the Kalman filter operates as an ``inner loop'' within a broader ``outer loop'' that explores the full parameter space of the model.  This outer loop samples over the kernel parameters $\tau_i$ and $\sigma_i$ for each station, as well as over the visibility phases $\phi_{ij}$ for each baseline.  For each sample of $\tau_i$, $\sigma_i$, and $\phi_{ij}$, the Kalman filter inner loop provides the marginal likelihood -- i.e., the likelihood of the data (given these model parameters) after marginalizing over the gain phase parameters -- specified in \autoref{eqn:MarginalLikelihood}.  This marginal likelihood value is then combined (via Bayes' Theorem) with prior information for each parameter to update the sampler. The priors for our model parameters are as follows: 
\begin{itemize}
    \item The variation timescale $\tau_i$ for each station follows a truncated normal distribution with mean 0\,s and standard deviation 30\,s; the distribution is truncated at zero such that negative values are not permitted.
    \item The magnitude of the phase fluctuations $\sigma_i$ for each station follows a truncated normal distribution with a mean of 0 radians and a standard deviation of 2 radians; the distribution is truncated at zero such that negative values are not permitted.
    \item The visibility phases $\phi_{ij}$ follow a uniform prior over the range ($-4\pi$, $4\pi$) radians.
\end{itemize}

% For the analyses presented in this paper, we use the \texttt{dynesty} nested sampler to explore the model parameter space and to produce posterior distributions \citep{dynesty}.\footnote{\texttt{Dynesty.jl}, a Julia wrapper for \texttt{dynesty}, is available at \url{https://github.com/tkoolen/Dynesty.jl}.}

For the analyses presented in this paper, we use \texttt{dynesty} to explore the model parameter space and to produce posterior distributions \citep{dynesty}.\footnote{\texttt{Dynesty.jl}, a Julia wrapper for \texttt{dynesty}, is available at \url{https://github.com/tkoolen/Dynesty.jl}.}  The \texttt{dynesty} code implements the nested sampling algorithm \citep{Skilling_2004,Skilling_2006}, which was originally developed to evaluate Bayesian evidence integrals but which has since been broadly applied as a posterior estimation tool.

\section{Demonstration using synthetic data} \label{sec:ApplicationSynthetic}

\begin{table}[t]
\centering
\caption{Input values for synthetic data generation}
    \begin{tabular}{l c c c} 
     \toprule
     \textbf{Station/} & \textbf{\( \tau \)} & \textbf{\( \sigma^2 \)} & \textbf{\( \phi \)} \\ 
     \textbf{Baseline} & \textbf{(Timescale)} & \textbf{(Variance)} & \textbf{(Phase)} \\ 
     \midrule
     1 & 20 & 1.0 & - \\ 
     2 & 25 & 2.0 & - \\ 
     3 & 30 & 1.5 & - \\ 
     4 & 35 & 0.5 & - \\ 
    \midrule
     12 & - & - & 1.0 \\ 
     13 & - & - & 0.5 \\ 
     14 & - & - & 2.0 \\ 
     23 & - & - & 1.5 \\ 
     24 & - & - & 0.0 \\ 
     34 & - & - & 1.0 \\ 
     \bottomrule
    \end{tabular}
    \label{tab:true_values}
    \end{table}
    
    \begin{figure*}[t]
    \centering
    \includegraphics[width=\textwidth]{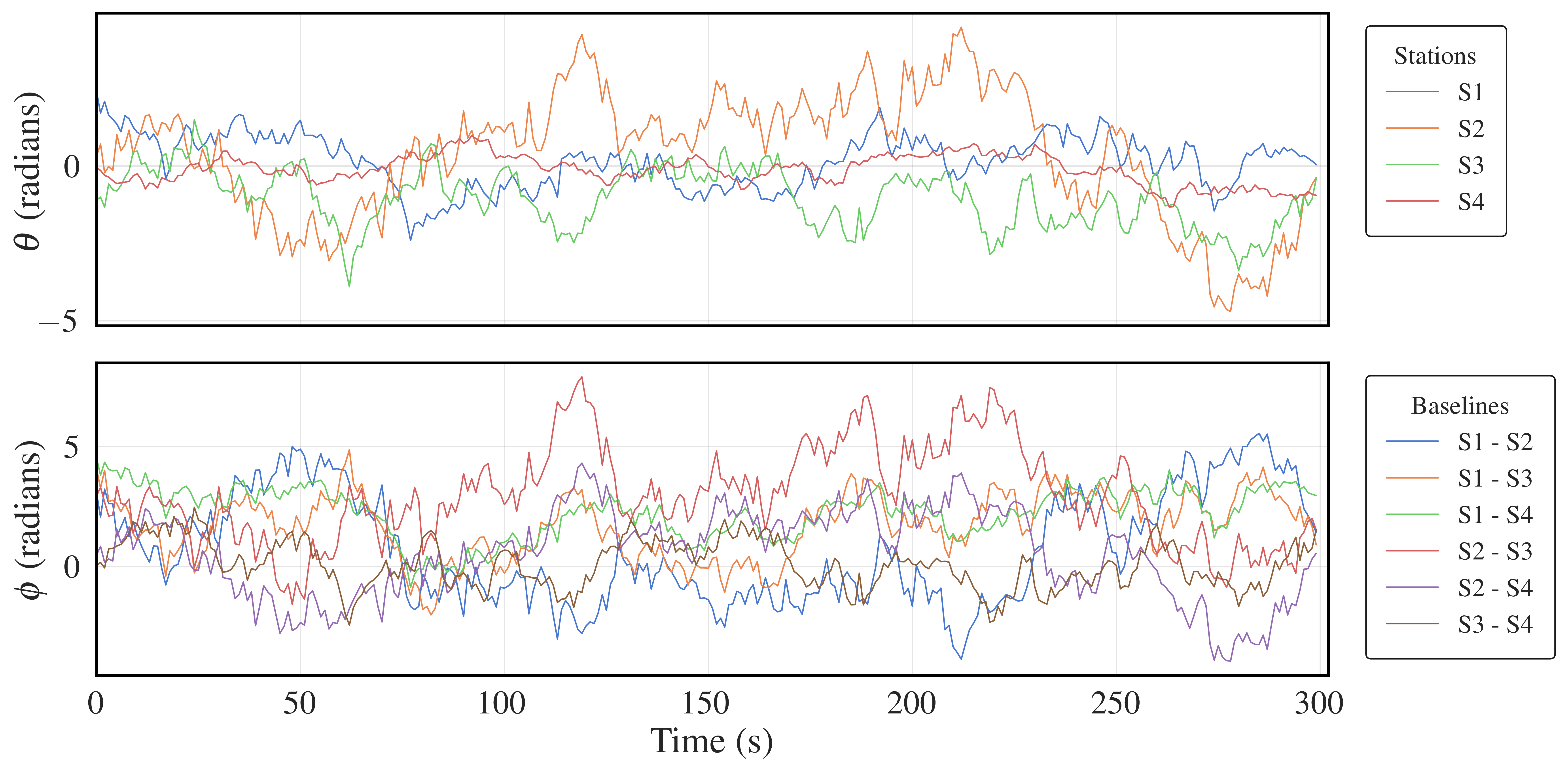}
    \caption{Synthetic data gain phases (top) and visibility phases (bottom) versus time for each station and baseline, respectively.
    }
    \label{fig:combined_gain_baseline}
    \end{figure*}
    
    In this section, we provide a demonstration of our phase modeling approach using synthetic data.  We generate synthetic visibility phase data appropriate for a 4-station array following the model described in \autoref{sec:Instrument}.
    
\begin{figure*}[t]
\centering
\includegraphics[width=\textwidth]{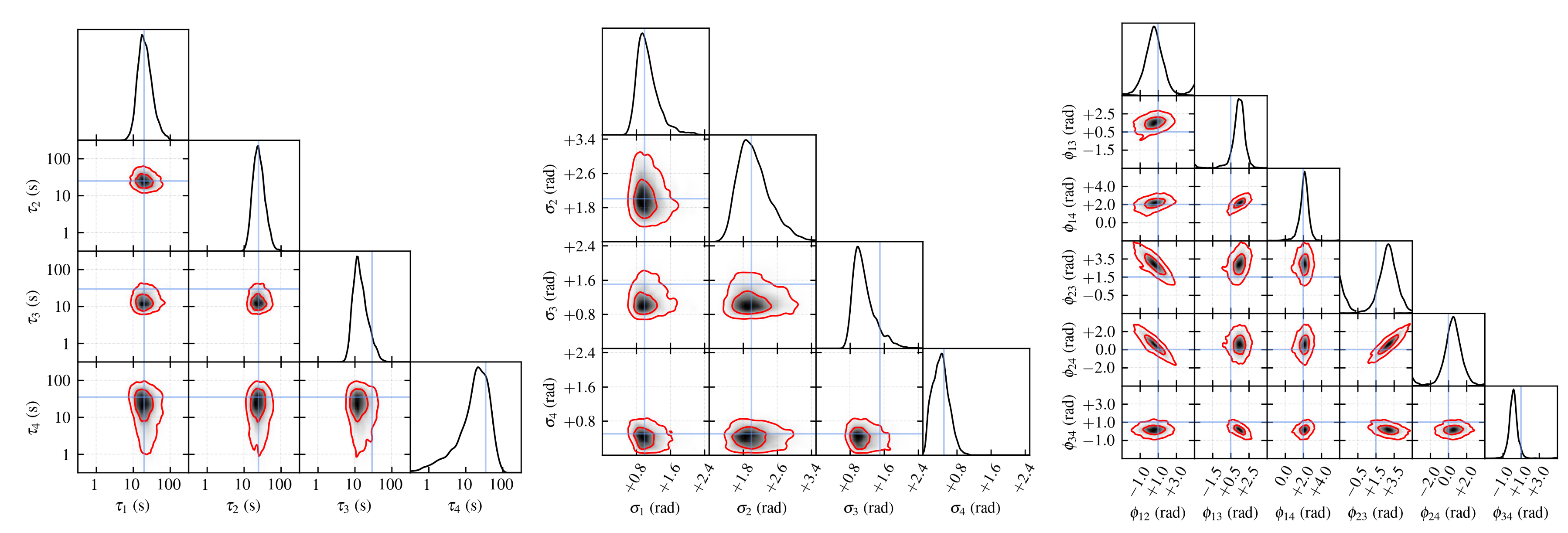}
\caption{Posterior distributions for the model parameters determined from fitting synthetic data. Left: Pairwise distributions for the coherence timescale parameters $\tau_i$. Center: Pairwise distributions for the phase fluctuation magnitude parameters $\sigma_i$. Right: Posterior distributions for the visibility phase parameters $\phi_{ij}$. For all 2D distributions, the red contours enclose 50\% and 90\% of the posterior probability. Blue lines represent the input values used when generating the synthetic data.}
\label{fig:synthetic_pairplots}
\end{figure*}

    For each station, we generate a time series of 300 gain phases via a random draw from a GP with $\tau_i$ and $\sigma_i$ parameters as specified in \autoref{tab:true_values}. The parameters for the synthetic datasets aim to reflect typical values observed in high-frequency VLBI. The $\tau$ values of $\sim$tens of seconds represent common atmospheric coherence times at 230 GHz \citep{EHT_M87_Paper2}, and the $\sigma$ values ($\sim 1$ rad) generate total phase fluctuations that are qualitatively similar to those in the 3C 279 data shown in   \autoref{fig:synthetic_pairplots}.  We use the \texttt{KernelFunctions.jl}\footnote{\url{https://juliagaussianprocesses.github.io/KernelFunctions.jl}} Julia package to sample the GPs, and we assume a Mat\'ern-1/2 kernel for all stations (see \autoref{eqn:Matern12}).  Each baseline in the array is assigned an intrinsic $\phi_{ij}$ visibility phase (specified in \autoref{tab:true_values}), whose value is modified by the gain phases per \autoref{eqn:Model}.  We then add to each timestamp on each baseline an independent random phase drawn from a Gaussian distribution with variance $\sigma_{ij}^2$, to simulate thermal noise. The value chosen for the thermal noise is 0.05. The generated gain phases for each station and the corresponding visibility phases on each baseline are shown in the top and bottom panels of \autoref{fig:combined_gain_baseline}, respectively.

After generating the synthetic data, we sample the posterior parameter space using the scheme described in \autoref{sec:ParameterExploration}.  \autoref{fig:synthetic_pairplots} shows the posterior distributions for the $\tau_i$ and $\sigma_i$ parameters describing the GP for each station, and the posterior distributions for the $\phi_{ij}$ visibility phase values.  We find accurate recovery of all model parameters.

\section{Application to real data} \label{sec:ApplicationReal}

\begin{figure*}[t]
\centering
\includegraphics[width=\textwidth]{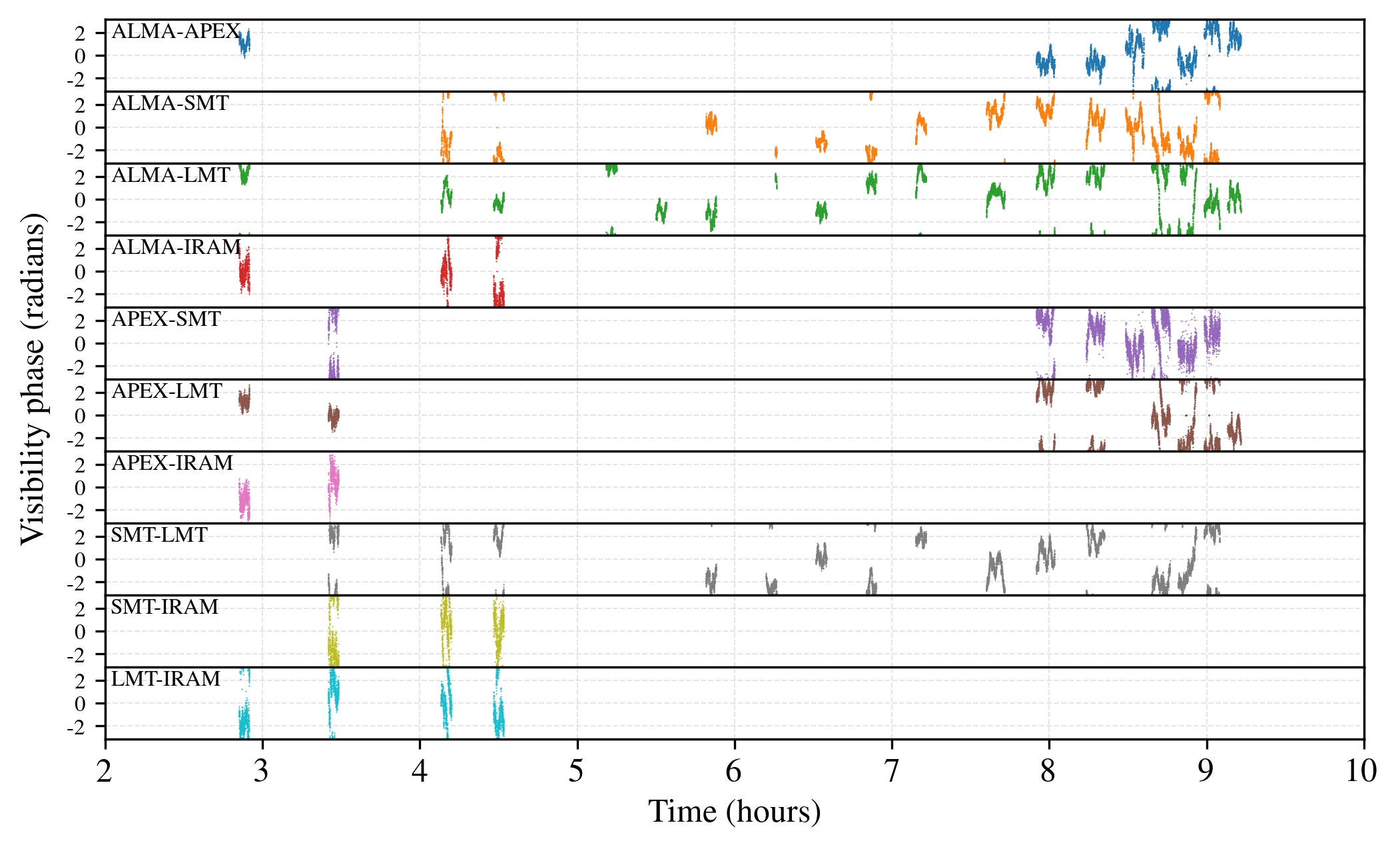}
\caption{Uncalibrated visibility phase versus time on each of the baselines in the EHT 3C\,279 dataset; in each panel, the baseline is specified in the upper left-hand corner.  Time is measured in UT hours since the start of 2017 April 5.}
\label{fig:realdata_allscans}
\end{figure*}

\begin{figure}[t]
\centering
\includegraphics[width=\columnwidth]{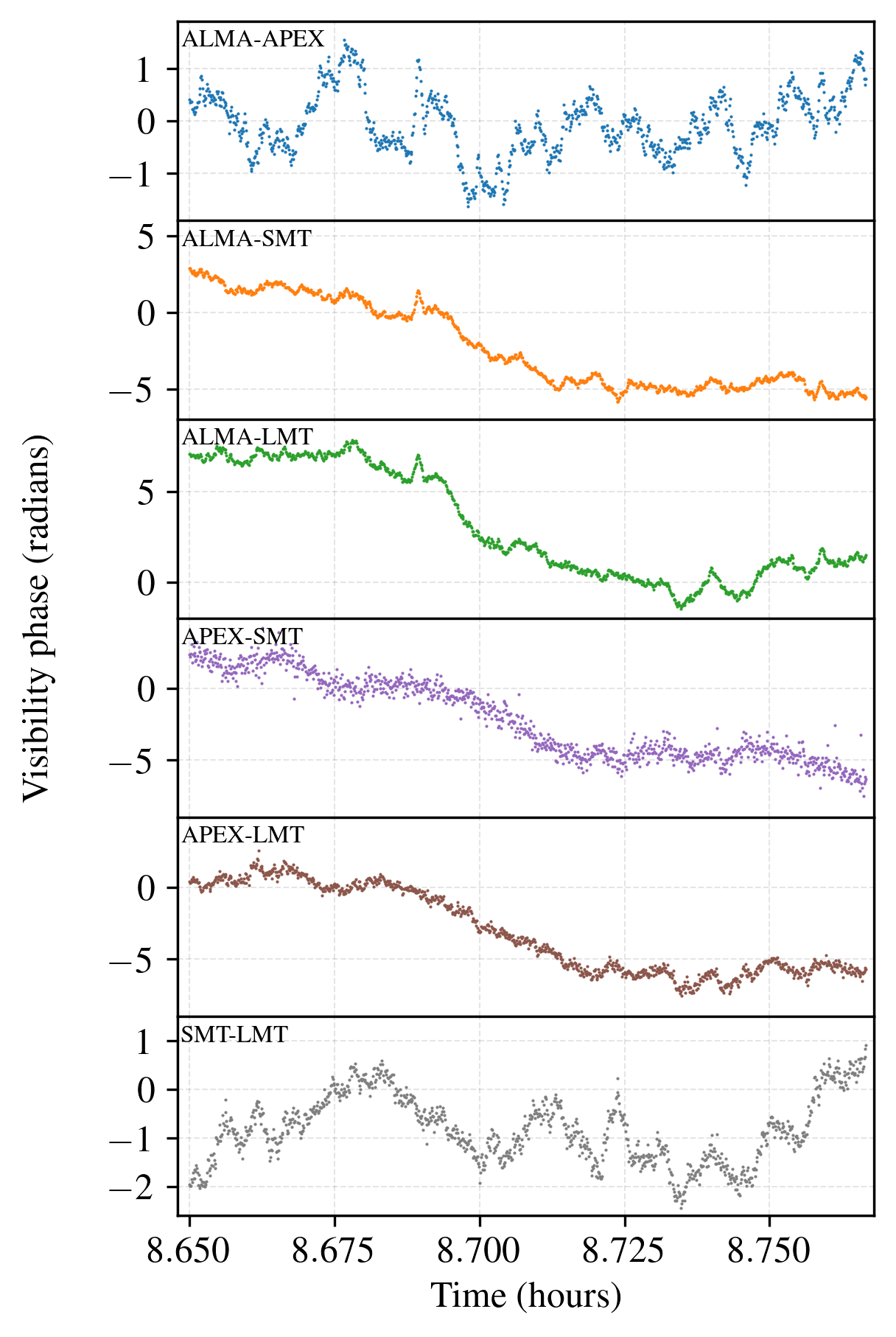}
\caption{Same as \autoref{fig:realdata_allscans}, but showing now the unwrapped visibility phases for a single scan.  Note that no single scan contained all 10 baselines shown in \autoref{fig:realdata_allscans}, so we show here a scan containing only 6 baselines.}
\label{fig:realdata_singlescan}
\end{figure}

\begin{figure}[t]
\centering
\includegraphics[width=\columnwidth]{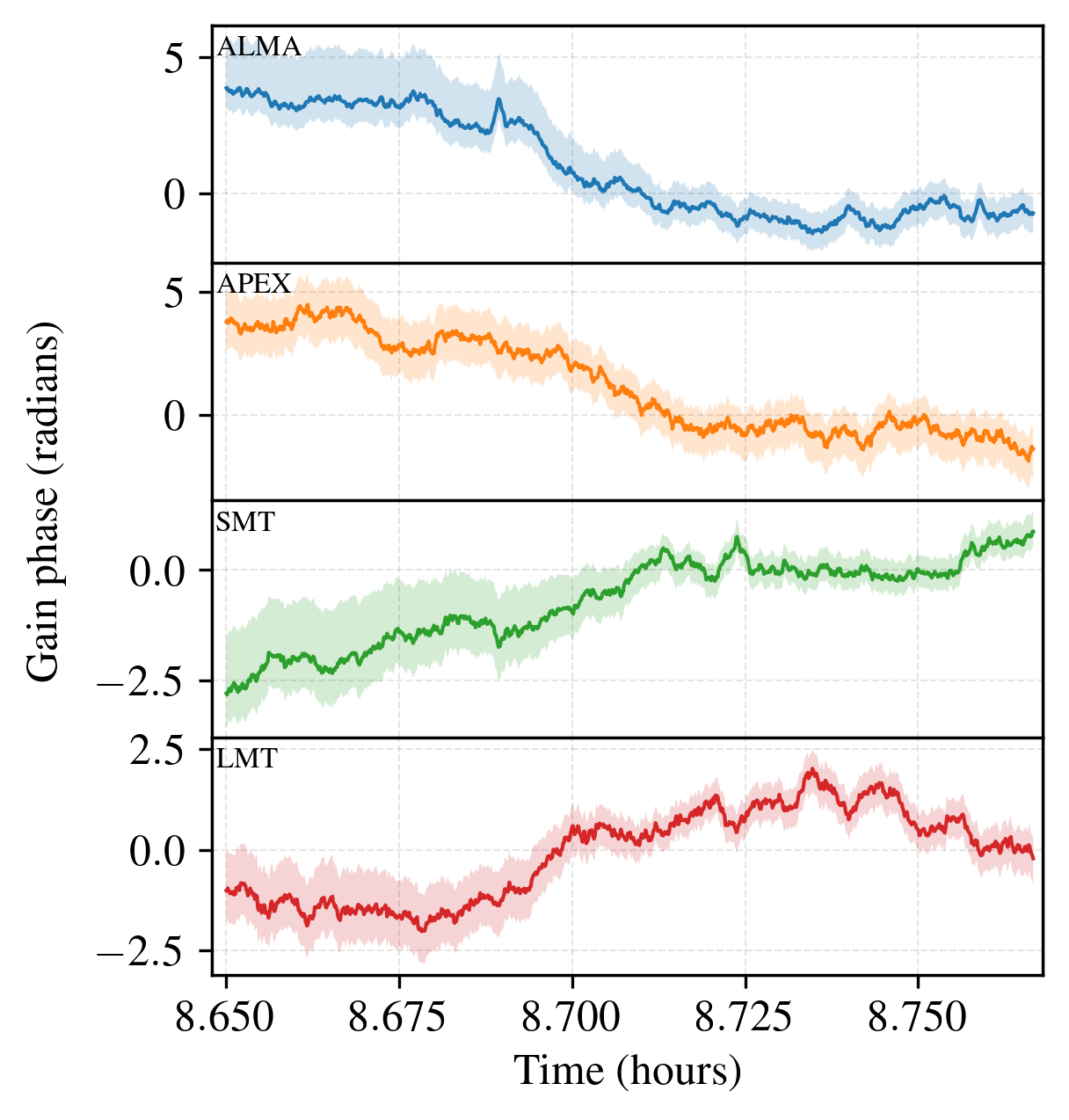 }
\caption{Recovered station gain phases versus time for each of the stations participating in the scan whose visibility phases are shown in \autoref{fig:realdata_singlescan}.  The solid curve in each panel shows the median posterior value, and the shaded region indicates 90\% confidence interval.
}
\label{fig:gain_fits_scan15}
\end{figure}

\begin{figure}[t]
\centering
\includegraphics[width=\columnwidth]{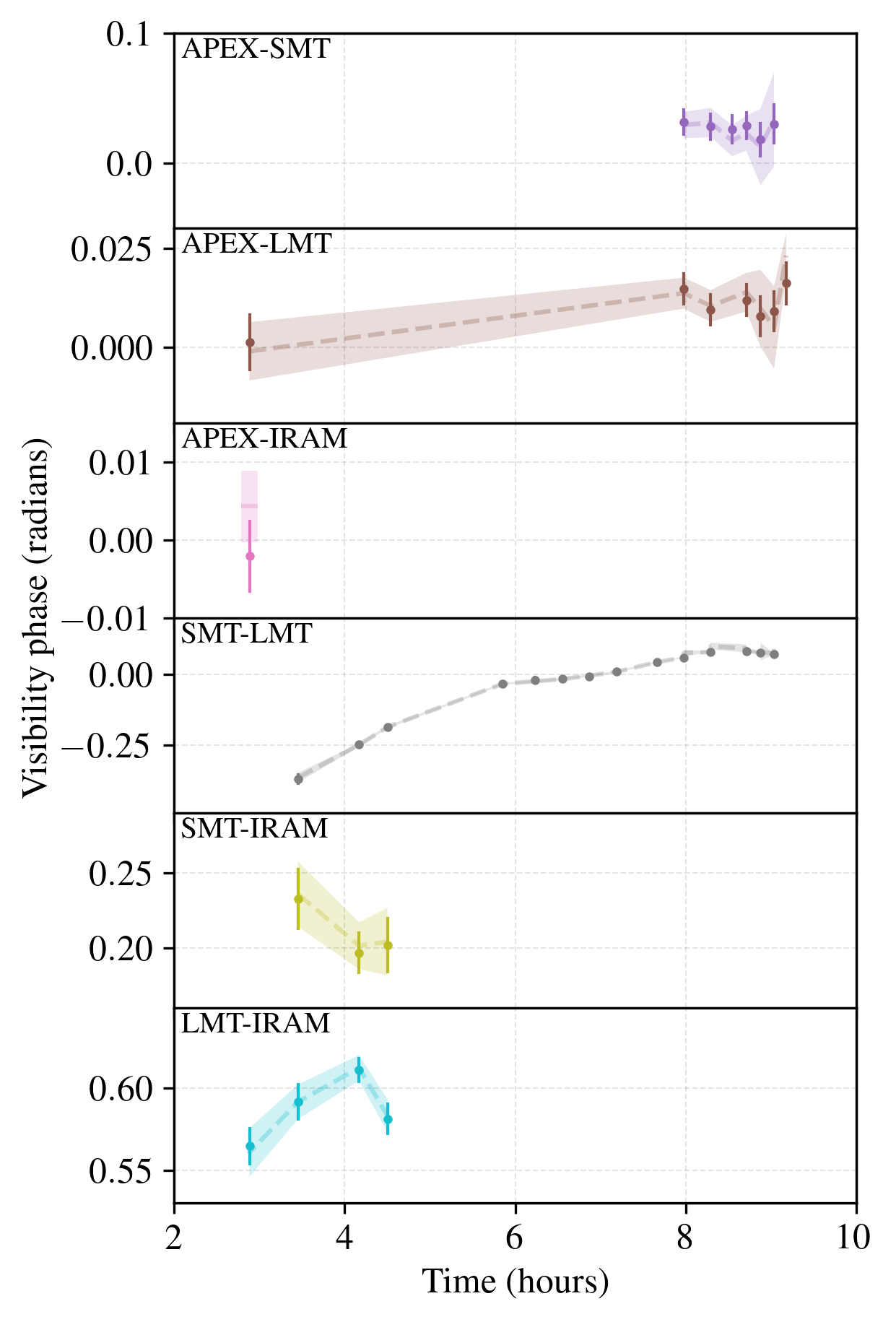}
\caption{Calibrated and scan-averaged visibility phase versus time on each of the non-ALMA baselines in the EHT 3C\,279 dataset; in each panel, the baseline is specified in the upper left-hand corner.  The data calibrated by the EHT Collaboration are shown as dashed lines with shading indicating $1\sigma$ confidence regions, while the results of the calibration carried out in this paper are shown as circular markers with error bars (again indicating $1\sigma$ confidence regions).}
\label{fig:visibility_phases_fitted}
\end{figure}

In this section, we apply our phase model to real high-frequency VLBI data collected by the EHT.  We use a dataset corresponding to an EHT observation of the blazar 3C\,279 carried out on 2017 April 5.  This blazar is a bright calibrator source with many high signal-to-noise ratio detections in the EHT data, which are important for reliably unwrapping visibility phases.  Details about the original processing and calibration of this dataset can be found in \cite{EHT_M87_Paper3}, and the scientific utility has been presented in \cite{Kim_2020}.

For our analysis, we work with a version of the original EHT dataset that has been partially calibrated using the \texttt{EHT-HOPS} pipeline \citep{Blackburn_2019}.  Specifically, we only carry out calibration through ``stage 2'' of the typical \texttt{EHT-HOPS} pipeline, such that phase bandpass solutions have been applied but no time-dependent phase calibration has yet taken place beyond a single delay-rate correction per scan.  We coherently average the data in frequency across the entire 2\,GHz band, but we retain the original post-correlation time segmentation of 0.4\,s.  For the proof-of-concept demonstration presented in this paper, we work with only the ``low-band'' data (centered on an observing frequency of 227\,GHz) and only with the left circularly polarized parallel-hand correlation products (LL).  We further restrict our analysis to just 5 of the 7 stations originally participating in the observation -- retaining ALMA, APEX, LMT, SMT, and the IRAM 30\,m but flagging SMA and JCMT -- so that the signal-to-noise ratio of the remaining data are high enough to disambiguate phase wraps at the native time sampling of 0.4\,s.\footnote{Proper handling of the phase unwrapping for the baselines to the SMA and JCMT stations -- which are located on Mauna Kea in Hawaii, and thus only participate in long baselines that have a generically lower SNR -- would require using a different averaging timescale compared to other baselines, which complicates the analysis without adding any insight.  So while the baselines contributed by these stations would be useful for, e.g., an imaging analysis, for the purposes of the calibration demonstration presented here, we have opted instead to stick with a more uniform subset of high-SNR baselines.}  We unwrap the visibility phases prior to fitting using the \texttt{numpy} library's \texttt{unwrap} function.  \autoref{fig:realdata_allscans} shows the visibility phases versus time for all scans in this dataset, where individual scans can be identified as groups of data points separated by visible gaps in time.  \autoref{fig:realdata_singlescan} shows the unwrapped visibility phases versus time for a single scan, and \autoref{fig:gain_fits_scan15} shows our gain phase solution for that same scan.

The 3C\,279 dataset is divided into 19 individual scans, each of which has a duration of several minutes.  We expect the constant-valued visibility phase assumption to hold for each scan individually, but not across many scans, so we independently fit our model to each individual scan.  The resulting calibrated visibility phases are shown in \autoref{fig:visibility_phases_fitted}, where they are compared against the scan-averaged output of the full \texttt{EHT-HOPS} calibration from \cite{EHT_M87_Paper3}.\footnote{We note that the data calibrated with \texttt{EHT-HOPS} are phase-referenced to the ALMA station, such that visibility phases on baselines to ALMA are zero-valued while retaining phase closure.  For the comparison in \autoref{fig:visibility_phases_fitted}, we thus carry out a post-fitting phase referencing to the ALMA station, though we note that no such referencing is required during the fitting procedure itself.}  We find quantitative agreement between the results of our fitting routine and the fully-calibrated EHT data, both in the values of the visibility phases and in the magnitudes of their statistical uncertainties.

\section{Discussion}\label{sec:Discussion}

Our phase model and associated fitting routine provides a new tool for calibrating visibility phase measurements.  The choice of a GP prescription for the individual station gains affords the model substantial flexibility in describing phase variations, which is necessary for capturing the rapid phase fluctuations typically introduced by the atmosphere during high-frequency VLBI observations.

For both the synthetic data and real data demonstrations presented in this paper, we have adopted a Mat\'ern-1/2 covariance kernel for the GPs describing each of the station gain phases.  This specific choice of kernel is not intrinsic to our phase model or our fitting procedure, and in principle many different kernels could be selected and employed in an analogous manner.  In \autoref{app:Kernel}, we assess the sensitivity of our parameter measurements to the particular choice of kernel function, demonstrating that for the primary parameters of interest -- i.e., the visibility phases -- the Mat\'ern-1/2 kernel is a sufficiently flexible choice. An additional benefit of the Mat\'ern family of kernels is that they permit use of the efficient Kalman filtering algorithm, which is not a property that is shared by all possible kernels \citep[see, e.g.,][]{Hartikainen_2010}.

We have implemented our model using the Julia scientific programming language, which will permit its straightforward incorporation into the \texttt{Comrade.jl} Bayesian modeling framework for radio interferometry \citep{Tiede_2022}.  Within the \texttt{Comrade.jl} framework, our phase model can serve as a modular component of a comprehensive instrument model, and the constant-valued visibility phase assumption (see \autoref{sec:Instrument}) can be relaxed in favor of visibility phase values that are self-consistently informed by a parameterized model (e.g., image) for the source structure.  Furthermore, forward-modeling of the gain phases in this manner permits straightforward generalizations, such as inclusion of the deterministic frequency dependence expected for non-dispersive tropospheric phase contributions \citep[e.g.,][]{Carilli_1999}; we expect such model extensions to be the subject of future development.

\subsection{Reference station considerations} \label{sec:ReferenceStation}

One useful feature of our approach is that it does away with the notion of a reference station, the specification of which is a typical requirement in other phase calibration methods to remove trivial phase degeneracies.

For a fully-connected array containing $N$ stations, there are $B = N(N-1)/2$ baseline-based visibility phase parameters and $N$ station-based gain phase parameters at each time (see \autoref{eqn:Model}).  However, there are only $B$ visibility phase measurements acting as constraints on the system, leaving an additional $N$ degrees of freedom unconstrained.  A standard approach for removing these additional degrees of freedom is to select one station -- the reference station -- and to fix (typically to zero) the values of all ($N-1$) visibility phases containing that station as well as the (single) gain phase for that station \citep[e.g.,][]{Blackburn_2019}.

There are a number of conceptual and practical drawbacks associated with the reference station approach.  Conceptually, the identification of a single station to receive special treatment requires an arbitrary choice to be made.  Furthermore, imposing fixed gain phase values for any single station effectively forces a calibration solution to redistribute any actual variations in that station's gain phases among the other stations in the array, potentially in conflict with the assumptions of a physical model for the phase gain behavior.  Selection of a reference station also runs into practical issues with real-world observations, where individual stations can join late or leave early; discontinuous phase jumps can result when the reference station changes partway through an observation.

In our approach, the GP specification of the gain phases acts to regularize the system and break the phase degeneracies in a global manner, without needing to select out a single station to be treated differently from the others.  For fixed GP parameters, unique solutions for the gain phases can be determined through, e.g., likelihood maximization (see \autoref{app:GainPhaseML}).  Even when fitting for the GP parameters, fixing the mean of each GP (which we set to zero) and imposing a finite-width prior on its variance (e.g., $\sigma_i^2$ in \autoref{eqn:Matern12}) is sufficient to break the phase degeneracies.

\subsection{Limitations}\label{sec:limitations}

We note that the current formulation of our model and fitting routine suffers from a number of limitations, primarily associated with our choice of visibility phase as the measured quantity of interest.  Interferometers directly measure complex visibilities, not phases, and the association of a unique phase value for any particular complex visibility measurement can only be made when the signal-to-noise ratio is sufficiently large.

For reliably ``unwrapping'' phases, we find that signal-to-noise ratios $\gtrsim$3 are often sufficient, but this threshold should be understood as a rule of thumb rather than a strict cutoff. The critical requirement is the ability to distinguish phases near $+180^\circ$ from those near $-180^\circ$. At SNR $\sim$1, the RMS phase variation between consecutive data points is approximately 1 radian, making it difficult to uniquely identify phase wraps. We expect the performance to degrade approximately continuously with decreasing SNR rather than exhibiting a sharp failure at some particular threshold.

% For reliably ``unwrapping'' phases, we find that a signal-to-noise ratios $\gtrsim$3 is often sufficient, but this threshold should be understood as a rule of thumb rather than a strict cutoff.
% ; our current model is thus only applicable to visibility data that have signal-to-noise ratios $\gtrsim$3, for which phases can be reliably ``unwrapped.''

Relatedly, our assumption in \autoref{eqn:ThermalNoise} that the thermal noise in the visibility phase measurements is Gaussian-distributed will also only hold in the limit of large signal-to-noise ratio \citep{Blackburn_2020}, meaning that for data with low or moderate signal-to-noise ratios the likelihood function we've adopted during parameter estimation is not statistically appropriate.  A generalization of our approach that directly models the complex visibilities \citep[which are themselves expected to be Gaussian-distributed in the complex plane; see][]{TMS} rather than just the visibility phases could plausibly overcome both of these limitations, though such an approach would also break the linearity of \autoref{eqn:Model} and thus require a more sophisticated parameter estimation routine than we have employed here.

\section{Summary and conclusions} \label{sec:Summary}

We present a model and associated parameter estimation routine for calibrating radio interferometric visibility phase measurements, appropriate for tackling the strong atmospheric phase variability typically impacting high-frequency VLBI observations.  We use a Gaussian Process (GP) prescription to model the individual station gain phases, which provides sufficient flexibility to capture even highly erratic phase behavior.  The GP prescription further permits use of the Kalman filtering algorithm for efficient gain marginalization, which we employ during parameter estimation.  By enforcing the GPs for all stations to have fixed mean and finite variance, our model establishes a global phase zeropoint and removes the need to specify an arbitrary reference station during calibration.

We have applied our model to both synthetic data and to a real high-frequency ($\sim$230\,GHz) VLBI dataset from the EHT. For the synthetic data, we demonstrate that the model faithfully recovers the input parameter values describing both the individual station GPs as well as the visibility phases (\autoref{fig:synthetic_pairplots}). For the real data, we find that our recovered visibility phase parameters are quantitatively consistent with the results of \texttt{EHT-HOPS} calibration carried out by the EHT Collaboration \citep{EHT_M87_Paper3}, both in the values of the phases and in the magnitudes of their uncertainties (\autoref{fig:visibility_phases_fitted}).  These applications demonstrate that our phase model provides a viable method for calibrating high-frequency VLBI data.

Looking ahead, one of the main development priorities for high-frequency VLBI projects -- including the EHT \citep{EHT_midrange}, the next-generation EHT \citep{Doeleman_2023}, and the Black Hole Explorer \citep{Johnson_2024} -- is the inclusion of multi-frequency observing capabilities.  A driving factor behind this push is a desire to leverage the ``frequency phase transfer'' (FPT) calibration technique \citep{Rioja_2023,Issaoun_2025}, whereby atmospheric phase solutions derived at a lower frequency are used to stabilize the phase and increase coherent integration times at a higher frequency.  Because the expected phase relationship between the two observing frequencies is known, an FPT calibration scheme is straightforward to implement within a forward-modeling framework such as the one we present here.

ur current model suffers from some limitations, discussed in \autoref{sec:limitations}.  Most notably, our method requires the ability to reliably unwrap visibility phases, which is typically possible only for SNR $\gtrsim$3. In practice, we expect that our phase model will serve most effectively as one calibration component within more comprehensive forward-modeling image reconstruction tools such as \texttt{ehtim} \citep{ehtim1,ehtim2} or \texttt{Comrade.jl} \citep{Tiede_2022}.  Future work will focus on generalizing our model from visibility phases to complex visibilities, and on expanding it to include frequency dependence of the phase signal.  We expect the resulting improvements to the quality and efficiency of interferometric phase calibration to afford corresponding benefits to downstream image reconstructions and associated scientific analyses.

\software{\texttt{ChatGPT} \citep{openai2022chatgpt}, \texttt{dynesty} \citep{dynesty}, \texttt{EHT-HOPS} \citep{Blackburn_2019}, \texttt{KernelFunctions.jl}, \texttt{matplotlib} \citep{Hunter_2007}, \texttt{numpy} \citep{numpy_2011,numpy_2020}}

\acknowledgments

We are grateful to Laurent Loinard for constructive comments that improved the quality of this paper.  We have used \texttt{ChatGPT} \citep{openai2022chatgpt} to assist with code and figure generation for this paper.

Support for this work was provided by the NSF through grants AST-1935980 and AST-2034306, and by the Gordon and Betty Moore Foundation through grants GBMF5278 and GBMF10423. This work has been supported in part by the Black Hole Initiative at Harvard University, which is funded by the John Templeton Foundation (grants 60477, 61479, and 62286) and the Gordon and Betty Moore Foundation (grant GBMF8273).

This paper makes use of the following ALMA data: 2016.1.01114.V, 2016.1.01176.V, 2016.1.01404.V, and 2016.1.01154.V. ALMA is a partnership of ESO (representing its member states), NSF (USA) and NINS (Japan), together with NRC (Canada), MOST and ASIAA (Taiwan), and KASI (Republic of Korea), in cooperation with the Republic of Chile. The Joint ALMA Observatory is operated by ESO, AUI/NRAO and NAOJ.

\bibliography{references}{}
\bibliographystyle{aasjournal}

\clearpage
\appendix
\numberwithin{equation}{section}

\section{Analytic expressions} \label{app:GainPhaseMarginalization}

Following \citet{Blackburn_2020}, we can compactly write \autoref{eqn:Model} for all measurements simultaneously using a vectorized notation,

\begin{equation}
\hat{\boldsymbol{\phi}} = \boldsymbol{\phi} + \boldsymbol{\Phi} \boldsymbol{\theta} + \boldsymbol{\varepsilon}. \label{eqn:ModelVectorized}
\end{equation}

\noindent Here, $\hat{\boldsymbol{\phi}}$ denotes a vector of measured visibility phases, $\boldsymbol{\phi}$ denotes a vector of modeled visibility phases, $\boldsymbol{\theta}$ denotes a vector of modeled gain phases, $\boldsymbol{\varepsilon}$ denotes a vector of the thermal noise values, and $\boldsymbol{\Phi}$ is a ``design matrix'' that maps the gain phases to the visibility phases.  The design matrix $\boldsymbol{\Phi}$ contains as its elements only values of $1$, $-1$, and $0$, and the specific construction of $\boldsymbol{\Phi}$ depends on how the $\theta_i$ and $\phi_{ij}$ parameters are ordered within the vectors $\boldsymbol{\theta}$ and $\boldsymbol{\phi}$, respectively.  For example, given a 3-station array with 2 timestamps of data, we could expand out \autoref{eqn:ModelVectorized} as

\begin{equation}
\begin{pmatrix}
\hat{\phi}_{12}(t_1) \\
\hat{\phi}_{12}(t_2) \\
\hat{\phi}_{13}(t_1) \\
\hat{\phi}_{13}(t_2) \\
\hat{\phi}_{23}(t_1) \\
\hat{\phi}_{23}(t_2)
\end{pmatrix} = \begin{pmatrix}
\phi_{12} \\
\phi_{12} \\
\phi_{13} \\
\phi_{13} \\
\phi_{23} \\
\phi_{23}
\end{pmatrix} + \begin{pmatrix}
1 & 0 & -1 & 0 & 0 & 0 \\
0 & 1 & 0 & -1 & 0 & 0 \\
1 & 0 & 0 & 0 & -1 & 0 \\
0 & 1 & 0 & 0 & 0 & -1 \\
0 & 0 & 1 & 0 & -1 & 0 \\
0 & 0 & 0 & 1 & 0 & -1
\end{pmatrix} \begin{pmatrix}
\theta_1(t_1) \\
\theta_1(t_2) \\
\theta_2(t_1) \\
\theta_2(t_2) \\
\theta_3(t_1) \\
\theta_3(t_2)
\end{pmatrix} + \begin{pmatrix}
\varepsilon_{12}(t_1) \\
\varepsilon_{12}(t_2) \\
\varepsilon_{13}(t_1) \\
\varepsilon_{13}(t_2) \\
\varepsilon_{23}(t_1) \\
\varepsilon_{23}(t_2)
\end{pmatrix} , \label{eqn:ModelVectorizedExample}
\end{equation}

\noindent where the design matrix appears as the $6 \times 6$ square matrix in the middle term on the right-hand side of the equation.  Note that because we assume that the $\phi_{ij}$ parameters are constant-valued in time, there are repeated elements in the first term on the right-hand side of \autoref{eqn:ModelVectorizedExample}.

We can similarly cast \autoref{eqn:ThermalNoise} as a multivariate Gaussian distribution,

\begin{equation}
\boldsymbol{\varepsilon} \sim \mathcal{N}(0, \textbf{S}) , \label{eqn:ThermalNoiseVectorized}
\end{equation}

\noindent where $\textbf{S}$ is a diagonal matrix containing as its entries the $\sigma_{ij}^2(t)$ thermal noise variances.  We can combine \autoref{eqn:ModelVectorized} and \autoref{eqn:ThermalNoiseVectorized} and rearrange to construct a likelihood function for the model parameters,

\begin{equation}
\mathcal{L}(\hat{\boldsymbol{\phi}} | \boldsymbol{\phi}, \boldsymbol{\tau}, \boldsymbol{\sigma}, \boldsymbol{\theta}) = \mathcal{N}_{\hat{\boldsymbol{\phi}}}(\boldsymbol{\phi} + \boldsymbol{\Phi} \boldsymbol{\theta}, \textbf{S}) . \label{eqn:LikelihoodFunction}
\end{equation}

\subsection{Gain phase marginalization}

Following the procedure detailed in Appendix C.5 of \citet{Blackburn_2020} and assuming uninformative (i.e., wide, flat) priors on each of the model parameters, we can apply Bayes' Theorem using this likelihood function and integrate over the $\boldsymbol{\theta}$ nuisance parameters to determine the marginalized posterior distribution over the remaining parameters,

\begin{equation}
p(\boldsymbol{\phi}, \boldsymbol{\tau}, \boldsymbol{\sigma} | \hat{\boldsymbol{\phi}}) = \mathcal{N}_{\boldsymbol{\phi}}(\hat{\boldsymbol{\phi}}, \boldsymbol{\Sigma}) . \label{eqn:MarginalPosterior}
\end{equation}

\noindent Here, the posterior covariance matrix $\boldsymbol{\Sigma}$ can be constructed as a block matrix in terms of the contributions from individual baselines.  For instance, for a three-station array we could write

\begin{equation}
\boldsymbol{\Sigma} = \begin{pmatrix}
\textbf{S}_{12} + \boldsymbol{\Sigma}_1 + \boldsymbol{\Sigma}_2 & \boldsymbol{\Sigma}_1 & - \boldsymbol{\Sigma}_2 \\
\boldsymbol{\Sigma}_1 & \textbf{S}_{13} + \boldsymbol{\Sigma}_1 + \boldsymbol{\Sigma}_3 & \boldsymbol{\Sigma}_3 \\
-\boldsymbol{\Sigma}_2 & \boldsymbol{\Sigma}_3 & \textbf{S}_{23} + \boldsymbol{\Sigma}_2 + \boldsymbol{\Sigma}_3
\end{pmatrix} ,
\end{equation}

\noindent where the $\textbf{S}_{ij}$ sub-matrices are diagonal and contain the as their entries the $\sigma_{ij}^2(t)$ thermal noise variances, and the $\boldsymbol{\Sigma}_i$ sub-matrices are the individual station GP covariances whose entries are given by \autoref{eqn:Matern12}.

Although \autoref{eqn:MarginalPosterior} is Gaussian, we note that the parameters describing the GP kernels for each station -- i.e., the $\tau_i$ and $\sigma_i$ parameters -- are not themselves Gaussian-distributed.  These parameters appear in the individual $\boldsymbol{\Sigma}_i$ sub-matrices, and are thereby subject to nonlinear operations (e.g., the determinant of $\boldsymbol{\Sigma}$) that result in a nontrivial posterior distribution.  Furthermore, $\boldsymbol{\Sigma}$ -- which is a $NT \times NT$ matrix -- has typical dimensions that are well in excess of $10^3$, making numerical sampling from \autoref{eqn:MarginalPosterior} extremely inefficient in practice.

\subsection{Maximum a posteriori visibility phase solution} \label{app:MAPvisphase}

Given fixed values of the GP kernel parameters $\tau_i$ and $\sigma_i$ for each station $i$, the maximum a posteriori (MAP) solution for $\boldsymbol{\phi}$ can be determined by setting all $\phi$-derivatives of \autoref{eqn:MarginalPosterior} (e.g., $\frac{\partial p}{\partial \phi_{12}}$, $\frac{\partial p}{\partial \phi_{13}}$, etc.) to zero.  The result can be expressed as

\begin{equation}
\boldsymbol{\phi}_{\text{MAP}} = \begin{pmatrix}
\boldsymbol{1}_{12}^{\top} \boldsymbol{\Sigma}^{-1} \boldsymbol{1}_{12} & \boldsymbol{1}_{12}^{\top} \boldsymbol{\Sigma}^{-1} \boldsymbol{1}_{13} & \ldots & \boldsymbol{1}_{12}^{\top} \boldsymbol{\Sigma}^{-1} \boldsymbol{1}_{N-1,N} \\
\boldsymbol{1}_{13}^{\top} \boldsymbol{\Sigma}^{-1} \boldsymbol{1}_{12} & \boldsymbol{1}_{13}^{\top} \boldsymbol{\Sigma}^{-1} \boldsymbol{1}_{13} & \ldots & \boldsymbol{1}_{13}^{\top} \boldsymbol{\Sigma}^{-1} \boldsymbol{1}_{N-1,N} \\
\vdots & \vdots & \ddots & \vdots \\
\boldsymbol{1}_{N-1,N}^{\top} \boldsymbol{\Sigma}^{-1} \boldsymbol{1}_{12} & \boldsymbol{1}_{N-1,N}^{\top} \boldsymbol{\Sigma}^{-1} \boldsymbol{1}_{13} & \ldots & \boldsymbol{1}_{N-1,N}^{\top} \boldsymbol{\Sigma}^{-1} \boldsymbol{1}_{N-1,N}
\end{pmatrix}^{-1} \begin{pmatrix}
\boldsymbol{1}_{12}^{\top} \boldsymbol{\Sigma}^{-1} \hat{\boldsymbol{\phi}} \\
\boldsymbol{1}_{13}^{\top} \boldsymbol{\Sigma}^{-1} \hat{\boldsymbol{\phi}} \\
\vdots \\
\boldsymbol{1}_{N-1,N}^{\top} \boldsymbol{\Sigma}^{-1} \hat{\boldsymbol{\phi}}
\end{pmatrix} ,
\end{equation}

\noindent where $\boldsymbol{\Sigma}$ is the posterior covariance (\autoref{eqn:MarginalPosterior}), and the notation $\boldsymbol{1}_{ij}$ indicates a vector whose entries are $1$ wherever $\boldsymbol{\phi}$ is equal to $\phi_{ij}$ and zero everywhere else.

\subsection{Maximum likelihood gain phase solution} \label{app:GainPhaseML}

Rather than marginalizing the gain phases, we can alternatively ask what the maximum-likelihood solution for $\boldsymbol{\theta}$ is, given some guess or solution for the other model parameters (i.e., the visibility phases $\boldsymbol{\phi}$ and the GP kernel parameters $\tau_i$, $\sigma_i$ for each station).  To identify a unique solution in the face of phase degeneracies (e.g., associated with an arbitrary phase zeropoint; see discussion in \autoref{sec:ReferenceStation}), the likelihood function \autoref{eqn:LikelihoodFunction} needs to be augmented with a prior; in our case, the gain phase prior is determined by the GPs at each station.  The effective likelihood function is then given by

\begin{equation}
\mathcal{L}_{\text{eff}} = \mathcal{N}_{\hat{\boldsymbol{\phi}}}(\boldsymbol{\phi} + \boldsymbol{\Phi} \boldsymbol{\theta}, \textbf{S}) \mathcal{N}_{\boldsymbol{\theta}}\left( 0 , \boldsymbol{\Sigma}_{\theta} \right) , \label{eqn:EffectiveLikelihood}
\end{equation}

\noindent where $\boldsymbol{\Sigma}_{\theta}$ encodes the covariance structure for all stations.  For instance, $\boldsymbol{\Sigma}_{\theta}$ for a three-station array could be constructed as a block-diagonal matrix,

\begin{equation*}
\boldsymbol{\Sigma}_{\theta} = \begin{pmatrix}
\boldsymbol{\Sigma}_1 & \boldsymbol{0} & \boldsymbol{0} \\
\boldsymbol{0} & \boldsymbol{\Sigma}_2 & \boldsymbol{0} \\
\boldsymbol{0} & \boldsymbol{0} & \boldsymbol{\Sigma}_3
\end{pmatrix} ,
\end{equation*}

\noindent where $\boldsymbol{\Sigma}_i$ is the covariance matrix describing the GP gain phase structure for station $i$ (see \autoref{eqn:GPphases} and \autoref{eqn:Matern12}) and $\boldsymbol{0}$ indicates a matrix whose entries are all zero.

As in \autoref{app:MAPvisphase}, the maximum-likelihood solution for $\boldsymbol{\theta}$ can be determined by setting all $\theta$-derivatives of \autoref{eqn:EffectiveLikelihood} to zero.  The result is given by

\begin{equation}
\boldsymbol{\theta} = \left( \boldsymbol{\Phi}^{\top} \textbf{S}^{-1} \boldsymbol{\Phi} + \boldsymbol{\Sigma}_{\theta}^{-1} \right)^{-1} \boldsymbol{\Phi}^{\top} \textbf{S}^{-1} \left( \hat{\boldsymbol{\phi}} - \boldsymbol{\phi} \right) ,
\end{equation}

\noindent which can also be understood as a standard generalized least squares solution to the system of linear equations described by \autoref{eqn:ModelVectorized} (plus terms associated with the prior).

\section{Kalman filter implementation} \label{app:KalmanFilters}

In this appendix, we provide a description of the Kalman filter implementation we use to estimate atmospheric gain phases.  We begin by establishing the model formulation, and we then describe the filter's prediction and update steps.

\subsection{Parameters and measurements}

For a VLBI array with N stations, we define the ``state vector'' $\boldsymbol{\theta}_k$ at time $t_k$ to be the collection of gain phases at each station,

\begin{equation}
\boldsymbol{\theta}_k = \begin{pmatrix}
\theta_{1}(t_k) \\
\theta_{2}(t_k) \\
\vdots \\
\theta_{N}(t_k)
\end{pmatrix} . \label{eqn:StateVector}
\end{equation}

\noindent The ``measurement vector'' $\tilde{\boldsymbol{\phi}}_k$ consists of the residual phases on each baseline after subtracting the model visibility phases,

\begin{equation}
\tilde{\boldsymbol{\phi}}_k = \begin{pmatrix}
\hat{\phi}_{12}(t_k) - \phi_{12}(t_k) \\
\hat{\phi}_{13}(t_k) - \phi_{13}(t_k) \\
\vdots \\
\hat{\phi}_{N-1,N}(t_k) - \phi_{N-1,N}(t_k)
\end{pmatrix} .
\end{equation}

\noindent Our measurement model relates the measurement vector to the state vector,

\begin{equation}
p(\tilde{\boldsymbol{\phi}}_k | \boldsymbol{\theta}_k) = \mathcal{N}_{\tilde{\boldsymbol{\phi}}_k}\left( \boldsymbol{\Phi}_k \boldsymbol{\theta}_k , \textbf{S}_k \right) ,
\end{equation}

\noindent where $\boldsymbol{\Phi}_k$ is the design matrix mapping station phases to baseline phases at timestamp $t_k$ \citep{Blackburn_2020}, and $\textbf{S}_k$ is a diagonal matrix containing as its nonzero elements the thermal noise variances for each baseline.

\subsection{Necessary conditions for Kalman filtering}

The Kalman filtering algorithm is only applicable to linear Gaussian “state-space” models. In the context of our phase model, this state-space requirement is equivalent to the assumption that the gain phases are a Markov process; i.e., the set of gain phases at time \(t_k\) must depend only on the set of gain phases at time \(t_{k-1}\). Because the measurements at each timestamp are independent (i.e., the measurement noise is assumed to be uncorrelated in time), we can write the general posterior conditioned on the observations at \(\{ t_1, t_2, \ldots, t_T \}\) as

\begin{equation}
p(\boldsymbol{\theta}_{1:T} | \tilde{\boldsymbol{\phi}}_{1:T}) \propto p(\boldsymbol{\theta}_{1:T}) \prod_{k=1}^T p_k(\tilde{\boldsymbol{\phi}}_k | \boldsymbol{\theta}_k) ,
\end{equation}

\noindent where we use the notation \(p_k(\tilde{\boldsymbol{\phi}}_k | \boldsymbol{\theta}_k) \equiv p(\tilde{\boldsymbol{\phi}}_k | \boldsymbol{\theta}_k)\) to indicate that the likelihood is specific to the measurement at time \(t_k\). The Markovian condition permits us to factorize the prior \(p(\boldsymbol{\theta}_{1:T})\) as

\begin{align}
p(\boldsymbol{\theta}_{1:T}) &= p(\boldsymbol{\theta}_T | \boldsymbol{\theta}_{1:T-1})\, p(\boldsymbol{\theta}_{1:T-1}) \nonumber \\ 
&= p(\boldsymbol{\theta}_T | \boldsymbol{\theta}_{T-1})\, p(\boldsymbol{\theta}_{1:T-1}) \nonumber \\
&= \prod_{k=1}^T p(\boldsymbol{\theta}_{k} | \boldsymbol{\theta}_{k-1}), \qquad \textit{(by recursion)} ,
\end{align}

\noindent where we let \(p(\boldsymbol{\theta}_1 | \boldsymbol{\theta}_0) = p(\boldsymbol{\theta}_1)\) for notational simplicity. The posterior can then be written as

\begin{equation}
p(\boldsymbol{\theta}_{1:T} | \tilde{\boldsymbol{\phi}}_{1:T}) = \prod_{k=1}^T p(\tilde{\boldsymbol{\phi}}_k | \boldsymbol{\theta}_{k})\, p(\boldsymbol{\theta}_k | \boldsymbol{\theta}_{k-1}) .
\end{equation}

\noindent With this formulation, one can show \citep[e.g.,][]{Sarkka_2013} that the marginal posterior at time \(t_k\) can be written as

\begin{equation}
p(\boldsymbol{\theta}_k | \tilde{\boldsymbol{\phi}}_{1:k}) = \frac{p(\tilde{\boldsymbol{\phi}}_k | \boldsymbol{\theta}_k)\, p(\boldsymbol{\theta}_k|\tilde{\boldsymbol{\phi}}_{1:k-1})}{\int d\boldsymbol{\theta}_k\, p(\tilde{\boldsymbol{\phi}}_k | \boldsymbol{\theta}_k)\, p(\boldsymbol{\theta}_k | \tilde{\boldsymbol{\phi}}_{1:k-1})} .
\end{equation}

\noindent This structure enables fast, sequential updates to the marginal posterior that “build up” the full posterior one timestamp at a time.

Not all possible GP kernels admit a state-space formulation, but it can be shown \citep{Rozanov_1977, rozanov1982markov} that all half-integer Mat\'ern kernels (e.g., Mat\'ern-1/2, Mat\'ern-3/2, etc.) do adhere to this condition.

\subsection{Transition distribution}

A key distribution in Kalman filtering is the ``transition distribution'' $p(\boldsymbol{\theta}_k | \boldsymbol{\theta}_{k-1})$, which in our case describes how the station gain phases evolve between two consecutive timestamps. Each gain phase $\theta_i$ evolves according to a Gaussian process specific to station $i$. Consequently, the covariance matrix $\boldsymbol{\Sigma}(t_n,t_m)$ for the joint distribution of station gains at times $t_n$ and $t_m$ can be constructed as a block matrix,

\begin{align}\label{eq:block}
\boldsymbol{\Sigma}(t_n,t_m) & = \begin{bmatrix}
\begin{pmatrix}
\Sigma_1(t_n,t_n) & 0 & \ldots & 0 \\
0 & \Sigma_2(t_n,t_n) & \ldots & 0 \\
\vdots & \vdots & \ddots & \vdots \\
0 & 0 & \ldots & \Sigma_N(t_n,t_n)
\end{pmatrix} & \begin{pmatrix}
\Sigma_1(t_n,t_m) & 0 & \ldots & 0 \\
0 & \Sigma_2(t_n,t_m) & \ldots & 0 \\
\vdots & \vdots & \ddots & \vdots \\
0 & 0 & \ldots & \Sigma_N(t_n,t_m)
\end{pmatrix} \\
\begin{pmatrix}
\Sigma_1(t_m,t_n) & 0 & \ldots & 0 \\
0 & \Sigma_2(t_m,t_n) & \ldots & 0 \\
\vdots & \vdots & \ddots & \vdots \\
0 & 0 & \ldots & \Sigma_N(t_m,t_n)
\end{pmatrix} & \begin{pmatrix}
\Sigma_1(t_m,t_m) & 0 & \ldots & 0 \\
0 & \Sigma_2(t_m,t_m) & \ldots & 0 \\
\vdots & \vdots & \ddots & \vdots \\
0 & 0 & \ldots & \Sigma_N(t_m,t_m)
\end{pmatrix}
\end{bmatrix} \nonumber \\
& \equiv \begin{bmatrix}
\boldsymbol{\Sigma}_{\boldsymbol{\theta}} & \textbf{T}_{nm} \\
\textbf{T}_{nm} & \boldsymbol{\Sigma}_{\boldsymbol{\theta}}
\end{bmatrix} ,
\end{align}

\noindent where in the second line we've introduced some shorthand notation and used the fact that $\Sigma_i(t_m,t_n) = \Sigma_i(t_n,t_m)$ and $\Sigma_i(t_n,t_n) = \Sigma_i(t_m,t_m)$.  For the Mat\'ern-1/2 kernel (see \autoref{eqn:Matern12}), the two sub-matrices are diagonal with elements

\begin{subequations}
\begin{align}
\{ \boldsymbol{\Sigma}_{\boldsymbol{\theta}} \}_{ii} & = \sigma_i^2 \label{eqn:JointCovarianceDiag} \\
\{ \textbf{T}_{nm} \}_{ii} & = \sigma_i^2 \exp\left(-\frac{|t_n - t_m|}{\tau_i}\right) .
\end{align}
\end{subequations}

\noindent The joint distribution for the gain phases on two consecutive timestamps $t_{k-1}$ and $t_k$ is then given by

\begin{equation}
\begin{bmatrix}
\boldsymbol{\theta}_k \\
\boldsymbol{\theta}_{k-1}
\end{bmatrix} \sim \mathcal{N}\Big( 0,\boldsymbol{\Sigma}(t_k,t_{k-1}) \Big) ,
\end{equation}

\noindent from which the transition distribution $p(\boldsymbol{\theta}_k | \boldsymbol{\theta}_{k-1})$ can be determined as the conditional distribution \citep[see, e.g.,][]{MatrixCookbook}

\begin{equation}
p(\boldsymbol{\theta}_k | \boldsymbol{\theta}_{k-1}) = \mathcal{N}_{\boldsymbol{\theta}_k}\left( \textbf{T}_{k,k-1} \boldsymbol{\Sigma}_{\boldsymbol{\theta}}^{-1} \boldsymbol{\theta}_{k-1}, \boldsymbol{\Sigma}_{\boldsymbol{\theta}} - \textbf{T}_{k,k-1} \boldsymbol{\Sigma}_{\boldsymbol{\theta}}^{-1} \textbf{T}_{k,k-1} \right) . \label{eqn:TransitionDistribution}
\end{equation}

\noindent This transition distribution can also be derived within the state-space model formulation, as in \citet{Hartikainen_2010}.

\subsection{Algorithmic implementation}

For our Kalman filter implementation, we follow the algorithmic approach outlined in \citet{Sarkka_2013}. Here we provide key mappings between the \citet{Sarkka_2013} notation (left) and that of our model (right):
\begin{subequations}
\begin{align}
\boldsymbol{x}_k \quad & \longrightarrow \quad \boldsymbol{\theta}_k \\
\boldsymbol{y}_k \quad & \longrightarrow \quad \tilde{\boldsymbol{\phi}}_k \\
\textbf{A}_{k} \quad & \longrightarrow \quad \textbf{T}_{k+1,k} \boldsymbol{\Sigma}_{\boldsymbol{\theta}}^{-1} \\
\textbf{Q}_{k} \quad & \longrightarrow \quad \boldsymbol{\Sigma}_{\boldsymbol{\theta}} - \textbf{T}_{k+1,k} \boldsymbol{\Sigma}_{\boldsymbol{\theta}}^{-1} \textbf{T}_{k+1,k} \\
\textbf{H}_{k} \quad & \longrightarrow \quad \boldsymbol{\Phi}_k \\
\textbf{R}_{k} \quad & \longrightarrow \quad \textbf{S}_{k} \\
\textbf{P}_0 \quad & \longrightarrow \quad \boldsymbol{\Sigma}_{\boldsymbol{\theta}} .
\end{align}
\end{subequations}

\noindent Given some set of kernel parameters $\left\{ \sigma_i, \tau_i \right\}$ describing the Gaussian processes at each of the stations, we initialize the gain phases $\boldsymbol{\theta}_0$ using a mean-zero Gaussian prior:
\begin{equation}
p(\boldsymbol{\theta}_0) = \mathcal{N}_{\boldsymbol{\theta}_0}(0, \boldsymbol{\Sigma}_{\boldsymbol{\theta}}) ,
\end{equation}
\noindent with $\boldsymbol{\Sigma}_{\boldsymbol{\theta}}$ as defined in \autoref{eqn:JointCovarianceDiag}.  For the first timestamp ($k=1$), we perform an initial prediction step:
\begin{subequations}
\begin{align}
\boldsymbol{m}_1^- & = 0 \\
\textbf{P}_1^- & = \boldsymbol{\Sigma}_{\boldsymbol{\theta}} .
\end{align}
\end{subequations}
\noindent This prediction step is followed by the first update step, which computes:
\begin{enumerate}
    \item the ``innovation vector,''
    \begin{equation}
    \boldsymbol{v}_1 = \boldsymbol{y}_1 - \textbf{H}_1 \textbf{m}_1^- = \tilde{\boldsymbol{\phi}}_1 ;
    \end{equation}
    \item the ``innovation covariance,''
    \begin{equation}
    \textbf{J}_1 = \textbf{H}_1 \textbf{P}_1^- \textbf{H}_1^{\top} + \textbf{R}_1 = \boldsymbol{\Phi}_1 \boldsymbol{\Sigma}_{\boldsymbol{\theta}} \boldsymbol{\Phi}_1^{\top} + \textbf{S}_1 ;
    \end{equation}
    \item the ``Kalman gain,''
    \begin{equation}
    \textbf{K}_1 = \textbf{P}_1^- \textbf{H}_1^{\top} \textbf{J}_1^{-1} = \boldsymbol{\Sigma}_{\boldsymbol{\theta}} \boldsymbol{\Phi}_1^{\top}(\boldsymbol{\Phi}_1 \boldsymbol{\Sigma}_{\boldsymbol{\theta}} \boldsymbol{\Phi}_1^{\top} + \textbf{S}_1)^{-1} ;
    \end{equation}
    \item the updated state estimate,
    \begin{equation}
    \boldsymbol{m}_1 = \boldsymbol{m}_1^- + \textbf{K}_1 \boldsymbol{v}_1 = \boldsymbol{\Sigma}_{\boldsymbol{\theta}} \boldsymbol{\Phi}_1^{\top}(\boldsymbol{\Phi}_1 \boldsymbol{\Sigma}_{\boldsymbol{\theta}} \boldsymbol{\Phi}_1^{\top} + \textbf{S}_1)^{-1} \tilde{\boldsymbol{\phi}}_1 ;
    \end{equation}
    \item and the updated state covariance,
    \begin{equation}
    \textbf{P}_1 = \textbf{P}_1^- - \textbf{K}_1 \textbf{J}_1 \textbf{K}_1^{\top} = \boldsymbol{\Sigma}_{\boldsymbol{\theta}} - \boldsymbol{\Sigma}_{\boldsymbol{\theta}} \boldsymbol{\Phi}_1^{\top}(\boldsymbol{\Phi}_1 \boldsymbol{\Sigma}_{\boldsymbol{\theta}} \boldsymbol{\Phi}_1^{\top} + \textbf{S}_1)^{-\top} \boldsymbol{\Phi}_1 \boldsymbol{\Sigma}_{\boldsymbol{\theta}}^{\top} .
    \end{equation}
\end{enumerate}

For subsequent timestamps ($k > 1$), we recursively apply the following steps:
\begin{subequations}
\begin{align}
\text{Predict:} \qquad \boldsymbol{m}_k^- & = \textbf{A}_{k-1} \boldsymbol{m}_{k-1} \\
\textbf{P}_k^- & = \textbf{A}_{k-1} \textbf{P}_{k-1} \textbf{A}_{k-1}^{\top} + \textbf{Q}_{k-1}
\end{align}
\end{subequations}
\begin{subequations}
\begin{align}
\text{Update:} \qquad \boldsymbol{v}_k & = \boldsymbol{y}_k - \textbf{H}_k \textbf{m}_k^- \\
\textbf{J}_k & = \textbf{H}_k \textbf{P}_k^- \textbf{H}_k^{\top} + \textbf{R}_k \\
\textbf{K}_k & = \textbf{P}_k^- \textbf{H}_k^{\top} \textbf{J}_k^{-1} \\
\boldsymbol{m}_k & = \boldsymbol{m}_k^- + \textbf{K}_k \boldsymbol{v}_k \\
\textbf{P}_k & = \textbf{P}_k^- - \textbf{K}_k \textbf{J}_k \textbf{K}_k^{\top} \\
\hphantom{\textbf{P}_k^-} & \hphantom{= \textbf{A}_{k-1} \textbf{P}_{k-1} \textbf{A}_{k-1}^{\top} + \textbf{Q}_{k-1}} \nonumber
\end{align}
\end{subequations}
\noindent This procedure yields the predictive distribution for the measurements at each time $t_k$ given all measurements at $t < t_k$,
\begin{equation}
p(\tilde{\boldsymbol{\phi}}_{k} | \tilde{\boldsymbol{\phi}}_{1:k-1}) = \mathcal{N}_{\tilde{\boldsymbol{\phi}}_{k}}(\textbf{H}_k \textbf{m}_k^-, \textbf{J}_k) .
\end{equation}
\noindent At the completion of filtering ($k = T$), we obtain the marginal likelihood,
\begin{equation}
p(\tilde{\boldsymbol{\phi}}_{1:T} | \boldsymbol{\tau}, \boldsymbol{\sigma}, \boldsymbol{\phi}) = \prod_{k=1}^T p(\tilde{\boldsymbol{\phi}}_{k} | \tilde{\boldsymbol{\phi}}_{1:k-1}) , \label{eqn:MarginalLikelihood}
\end{equation}
\noindent which is then used during parameter space exploration (see \autoref{sec:ParameterExploration}).

\section{Resilience to choice of kernel function} \label{app:Kernel}

In this section, we explore the resilience of our model to the choice of kernel function assumed by the Kalman filter.

As detailed in the main text and expanded upon in \autoref{app:KalmanFilters}, our implementation of the Kalman filter assumes a Mat\'ern-$1/2$ kernel.
However, we expect that real data will in general not adhere to this assumption.
To investigate the effects of this model choice, we generated synthetic datasets using the same approach described in \autoref{sec:ApplicationSynthetic}, but with the underlying kernel changed to Mat\'ern-$3/2$ and Mat\'ern-$5/2$.  
We continue to fit the data in all cases using a Kalman filter that assumes a  Mat\'ern-$1/2$ covariance structure.

For this experiment, we used the same input parameters as in \autoref{sec:ApplicationSynthetic} for $\tau$, $\sigma$, and $\phi$.

Because of the mismatch in model specification, we expect that the recovered parameter values will not in general match the input parameter values at the level nominally indicated by the posterior distribution.  However, of most interest for downstream use of the calibrated visibility phases is the fidelity with which the underlying astrophysical information is recovered.  We thus evaluate fit fidelity via comparison of the input and recovered closure phases $\psi_{ijk}$ on a triangle of baselines connecting stations $i$, $j$ and $k$, which are constructed from the visibility phases as

\begin{equation}
\psi_{ijk} = \phi_{ij} + \phi_{jk} + \phi_{ki} . \label{eq:close_phases}
\end{equation}

\noindent By construction, closure phases are independent of any station gain phases present in the data, and thus represent a ``robust'' observable quantity.

\begin{figure*}[t]
\centering
\includegraphics[width=\textwidth]{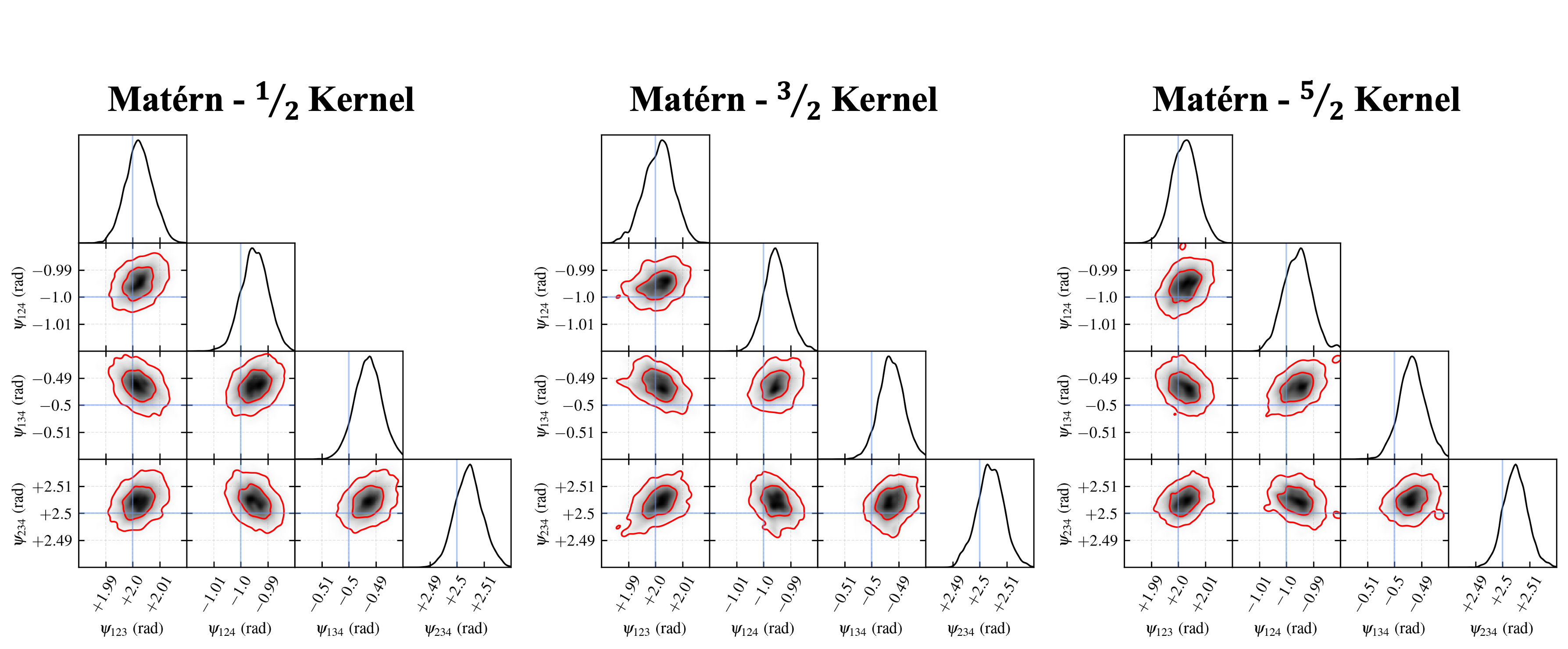}
\caption{Posterior distributions for the modeled closure phases ($\psi$), when fitting synthetic data generated using different GP kernels with a model that assumes the same kernel.  From left to right, the GP kernel used to generate the synthetic data is Mat\'ern-$1/2$, Mat\'ern-$3/2$, and Mat\'ern-$5/2$. Blue lines indicate the input $\psi$ values used to generate the data.}
\label{fig:closure_phases}
\end{figure*}

\autoref{fig:closure_phases} shows the recovered closure phase posterior distributions for each of our three test datasets.  We find that the recovered closure phases show good agreement with the input closure phases, even when the GP kernels are mismatched.

\end{document}